\documentclass[a4paper,conference]{IEEEtran}

\usepackage{cite}
\usepackage{amsmath,amssymb,amsfonts}
\usepackage{bbm}
\usepackage{algorithmic}
\usepackage{graphicx}
\usepackage{textcomp}
\usepackage{xcolor}
\usepackage{tabularx}
\usepackage{multirow} 				%
\usepackage{booktabs}
\usepackage[hidelinks,draft=false]{hyperref} %
\usepackage{siunitx}
\usepackage{tikz}
\usetikzlibrary{positioning}
\usepackage{pgf}
\usepackage{pgfplots}
\pgfplotsset{compat=1.16}
\usepgfplotslibrary{statistics}
\usepackage{orcidlink}
\definecolor{kit-green100}{rgb}{0,.59,.51}
\definecolor{kit-green70}{rgb}{.3,.71,.65}
\definecolor{kit-green50}{rgb}{.50,.79,.75}
\definecolor{kit-green30}{rgb}{.69,.87,.85}
\definecolor{kit-green15}{rgb}{.85,.93,.93}
\definecolor{KITgreen}{rgb}{0,.59,.51}

\definecolor{KITpalegreen}{RGB}{130,190,60}
\colorlet{kit-maigreen100}{KITpalegreen}
\colorlet{kit-maigreen70}{KITpalegreen!70}
\colorlet{kit-maigreen50}{KITpalegreen!50}
\colorlet{kit-maigreen30}{KITpalegreen!30}
\colorlet{kit-maigreen15}{KITpalegreen!15}

\definecolor{KITblue}{rgb}{.27,.39,.66}
\definecolor{kit-blue100}{rgb}{.27,.39,.67}
\definecolor{kit-blue70}{rgb}{.49,.57,.76}
\definecolor{kit-blue50}{rgb}{.64,.69,.83}
\definecolor{kit-blue30}{rgb}{.78,.82,.9}
\definecolor{kit-blue15}{rgb}{.89,.91,.95}

\definecolor{KITyellow}{rgb}{.98,.89,0}
\definecolor{kit-yellow100}{cmyk}{0,.05,1,0}
\definecolor{kit-yellow70}{cmyk}{0,.035,.7,0}
\definecolor{kit-yellow50}{cmyk}{0,.025,.5,0}
\definecolor{kit-yellow30}{cmyk}{0,.015,.3,0}
\definecolor{kit-yellow15}{cmyk}{0,.0075,.15,0}

\definecolor{KITorange}{rgb}{.87,.60,.10}
\definecolor{kit-orange100}{cmyk}{0,.45,1,0}
\definecolor{kit-orange70}{cmyk}{0,.315,.7,0}
\definecolor{kit-orange50}{cmyk}{0,.225,.5,0}
\definecolor{kit-orange30}{cmyk}{0,.135,.3,0}
\definecolor{kit-orange15}{cmyk}{0,.0675,.15,0}

\definecolor{KITred}{rgb}{.63,.13,.13}
\definecolor{kit-red100}{cmyk}{.25,1,1,0}
\definecolor{kit-red70}{cmyk}{.175,.7,.7,0}
\definecolor{kit-red50}{cmyk}{.125,.5,.5,0}
\definecolor{kit-red30}{cmyk}{.075,.3,.3,0}
\definecolor{kit-red15}{cmyk}{.0375,.15,.15,0}

\definecolor{KITpurple}{RGB}{160,0,120}
\colorlet{kit-purple100}{KITpurple}
\colorlet{kit-purple70}{KITpurple!70}
\colorlet{kit-purple50}{KITpurple!50}
\colorlet{kit-purple30}{KITpurple!30}
\colorlet{kit-purple15}{KITpurple!15}

\definecolor{KITcyanblue}{RGB}{80,170,230}
\colorlet{kit-cyanblue100}{KITcyanblue}
\colorlet{kit-cyanblue70}{KITcyanblue!70}
\colorlet{kit-cyanblue50}{KITcyanblue!50}
\colorlet{kit-cyanblue30}{KITcyanblue!30}
\colorlet{kit-cyanblue15}{KITcyanblue!15}

\usepackage{etoolbox}
\usepackage[printonly
used]{acronym}

\let\oldacrodef\acrodef
\let\oldac\ac
\let\oldAc\Ac
\let\oldacs\acs
\let\oldacl\acl
\let\oldAcl\Acl
\let\oldacp\acp
\let\oldAcp\Acp
\let\oldacsp\acsp

\renewcommand{\acrodef}[3][]{%
	\ifx&#1&%
	\oldacrodef{#2}[#2]{#3}%
	\newbool{used@#2}\setbool{used@#2}{false}
	\else
	\oldacrodef{#2}[#1]{#3}%
	\newbool{used@#1}\setbool{used@#1}{false}
	\fi
}

\renewcommand{\ac}[1]{%
	\ifbool{used@#1}{%
		\hyperref[#1:first]{\oldac{#1}}%
	}{%
		\oldac{#1}\label{#1:first}%
		\setbool{used@#1}{true}%
	}%
}

\renewcommand{\Ac}[1]{%
	\ifbool{used@#1}{%
		\hyperref[#1:first]{\oldAc{#1}}%
	}{%
		\oldAc{#1}\label{#1:first}%
		\setbool{used@#1}{true}%
	}%
}

\renewcommand{\acs}[1]{%
	\ifbool{used@#1}{%
		\hyperref[#1:first]{\oldacs{#1}}%
	}{%
		\oldacs{#1}\label{#1:first}%
		\setbool{used@#1}{true}%
	}%
}

\renewcommand{\acl}[1]{%
	\ifbool{used@#1}{%
		\hyperref[#1:first]{\oldacl{#1}}%
	}{%
		\oldacl{#1}\label{#1:first}%
		\setbool{used@#1}{true}%
	}%
}

\renewcommand{\Acl}[1]{%
	\ifbool{used@#1}{%
		\hyperref[#1:first]{\oldAcl{#1}}%
	}{%
		\oldAcl{#1}\label{#1:first}%
		\setbool{used@#1}{true}%
	}%
}

\renewcommand{\acp}[1]{%
	\ifbool{used@#1}{%
		\hyperref[#1:first]{\oldacp{#1}}%
	}{%
		\oldacp{#1}\label{#1:first}%
		\setbool{used@#1}{true}%
	}%
}

\renewcommand{\Acp}[1]{%
	\ifbool{used@#1}{%
		\hyperref[#1:first]{\oldAcp{#1}}%
	}{%
		\oldAcp{#1}\label{#1:first}%
		\setbool{used@#1}{true}%
	}%
}

\renewcommand{\acsp}[1]{%
	\ifbool{used@#1}{%
		\hyperref[#1:first]{\oldacsp{#1}}%
	}{%
		\oldacsp{#1}\label{#1:first}%
		\setbool{used@#1}{true}%
	}%
}

\definecolor{commentgreen}{rgb}{.1,.4,.1}
\usepackage{listings}
\lstdefinestyle{myshell}{
	backgroundcolor=\color{gray!30},   
	commentstyle=\color{green},
	keywordstyle=\color{magenta},
	numberstyle=\tiny\color{gray},
	stringstyle=\color{purple},
	basicstyle=\ttfamily\small,
	breakatwhitespace=false,         
	breaklines=true,                 
	captionpos=b,                    
	keepspaces=true,                                   
	showspaces=false,                
	showstringspaces=false,
	showtabs=false,                  
	tabsize=2,
	escapechar={|}, 
}

\newcommand\coloredtext[2][]{\tikz[baseline=-2pt]\node[minimum width=0.8em,minimum height=2.3ex, inner sep=0.1, fill=#1]{#2};}%

\newcommand{\us}{\rule[0pt]{5pt}{0.5pt}}
\newcolumntype{C}{>{\centering\arraybackslash}X}

\begin{document}

\acrodef{3GPP}{3rd Generation Partnership Project}
\acrodef{AI}{artificial intelligence}
\acrodef{BWP}{bandwidth part}
\acrodef{DCI}{downlink control information}
\acrodef{GELU}{Gaussian error linear unit}
\acrodef{gNB}{base station}
\acrodef{LLM}{large language model}
\acrodef{LSTM}{long short-term memory}
\acrodef{PDCCH}{physical downlink control channel}
\acrodef{PDSCH}{physical downlink shared channel}
\acrodef{PRB}{physical resource block}
\acrodef{PRE}{protocol reverse engineering}
\acrodef{PUCCH}{physical uplink control channel}
\acrodef{PUSCH}{physical uplink shared channel}
\acrodef{RAN}{radio access network}
\acrodef{RL}{relative Levenshtein}
\acrodef{RNTI}{radio network temporary identity}
\acrodef{RRC}{radio resource control}
\acrodef{SFN}{system frame number}
\acrodef{TP}{true positive}
\acrodef{UE}{user equipment}

\title{Physical Layer Message Prediction for 5G Radio Access Network Protocols}

\author{\IEEEauthorblockN{Jonathan Ebert\, \orcidlink{0009-0007-8613-7823} and Peter Rost\, \orcidlink{0000-0002-8341-6989}}
\IEEEauthorblockA{\textit{Communications Engineering Lab (CEL)} \\
\textit{Karlsruher Institute of Technology (KIT)}, Germany \\
\texttt{\{jonathan.ebert, peter.rost\}@kit.edu}}
}

\maketitle

\begin{abstract}
Protocol reverse engineering stands as the cutting-edge approach in security research. This paper presents a framework capable of reverse engineering the communications within a mobile communication system. Our focus is on systems released by the 3GPP, with an emphasis on 5G NR. 

Our approach leverages the available context and syntax of the 5G standard to predict subsequent messages. This approach relies on a Transformer model and is trained based on an open-source 5G system implementation, emulating a base station and several user equipments. The prediction targets messages at the physical layer.
\end{abstract}

\section{Introduction}
\subsection{Problem Description}

This research addresses protection against jamming and the enhancement of robustness in wireless communication protocols. In particular, we focus on the Fifth Generation (5G) mobile communication system as standardized by the \ac{3GPP}. Owing to the use of free-space wireless transmission in 5G, the system is inherently vulnerable to eavesdropping by potential adversaries. A key objective of this work is the prediction of future transmissions, as an attacker capable of accurately forecasting messages may be able to overshadow or disrupt them.

Transmission prediction in 5G presents significant challenges, as accurate forecasting requires consideration of the hidden internal states of the communicating entities. To address this, we develop a prediction framework based on \ac{PRE}, extended to incorporate user state information.

This paper specifically concentrates on prediction at the physical layer. The \ac{PRE} framework takes as input the messages exchanged between communication entities, focusing on control-plane messages, since user-plane data are encrypted and thus not accessible for analysis.

We focus on predicting messages at the physical layer because this capability can facilitate attacks such as intercepting downlink control information, e.g., shown by Ashik and Hossain \cite{ReaperPulse2025}. Alternatively, prediction can be used to distinguish legitimate transmissions from those that are being jammed, like what was demonstrated in \cite{Detecting5GSignal2024a}.

\subsection{Related Work}

State of the art \ac{PRE} utilizes unencrypted messages or the protocol’s implementation to extract information \cite{beddoe2004protocolInformatics,narayan_survey_2015}. \ac{PRE} mainly follows one of two strategies: the first involves deriving a state machine and associating messages with state transitions, while the second involves directly uncovering hidden patterns within message flows to forecast communication. 

One example of working directly with patterns in exchanged messages is the PREUNN method \cite{kiechle_preunn_2023}. This is an automated \ac{PRE} technique that leverages various machine learning models to reverse engineer unencrypted HTTP and FTP traffic. Their findings reveal that \ac{LSTM} models demonstrate considerable promise and successfully generate new traffic for system fuzzing. 

To infer the state machine, Wei et. al \cite{wei_inferring_2024} introduce a method to utilize \acp{LLM}. With targeted prompt engineering, they were able to extract state machines from the implementation of the protocols. In our case, the implementation is proprietary and beyond our reach. This led us to adopt Transformers, the core machine learning framework behind \acp{LLM}.

\subsection{Contribution}
We introduce a Transformer-based message predictor designed to infer future messages based on the contextual information and states of the communication scenario and its participants. Our system supports variable-length encrypted data and takes into account the time-frequency positions of messages as well as 5G-specific metadata. It also accommodates distinct traffic patterns across different \acp{UE}. We can forecast a slot in a 5G transmission scenario involving as many as 10 mobile phones, with good accuracy. This work should be seen as proof-of-concept since, to our knowledge, there exists no work we could compare with.
The prediction is made without any jammer present and is intended to represent the baseline performance.

\section{Preliminaries}
\subsection{5G Messages and Physical Channels}
The \ac{3GPP} specification outlines various channels at the physical layer. A channel in \ac{3GPP} refers to a section of the spectrum allocated for specific uses. The channels relevant to this study will be detailed in the next section. Further information can be found in the standard \cite{38.213,38.214}.

The uplink channels from \ac{UE} to \ac{gNB} are the \ac{PUSCH} and \ac{PUCCH}. The \ac{PUSCH} is used for uplink user data transmission, and the \ac{PUCCH} transmits channel state information used for rate adaptation, and power settings, to name a few use cases.

In the downlink direction, two channels are relevant to this work. The \ac{PDCCH} and the \ac{PDSCH}. The \ac{PDSCH} occupies most of the downlink resources and transports the user data and some control messages. The control channel conveys the unencrypted \ac{DCI}, which can be provided in different formats. The format 0\us0 and 0\us1 convey the uplink grant information, informing the \ac{UE} when it is allowed to send data on the \ac{PUSCH}. On the other side, a \ac{DCI} with format 1\us0 or 1\us1 indicates the position of data in the \ac{PDSCH} for a specific \ac{UE}. The \ac{RNTI} specifies the \acp{UE} that are associated with a particular \ac{PDCCH} message. The \ac{RNTI} is a unique identifier allocated by the 5G network to differentiate \acp{UE}.

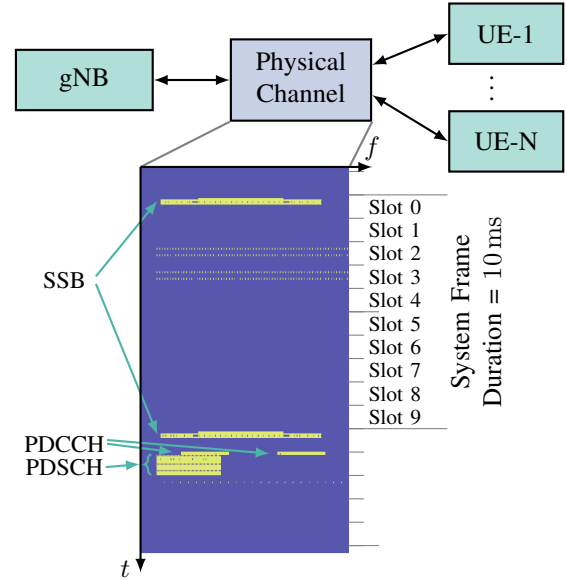
\begin{figure}[t!]
	\centering
	\begin{tikzpicture}[node distance=2cm, auto, >=latex, thick]
		\node[inner sep=0pt] (img) at (0,0) {\includegraphics[width=0.15\textwidth,height=5.1cm]{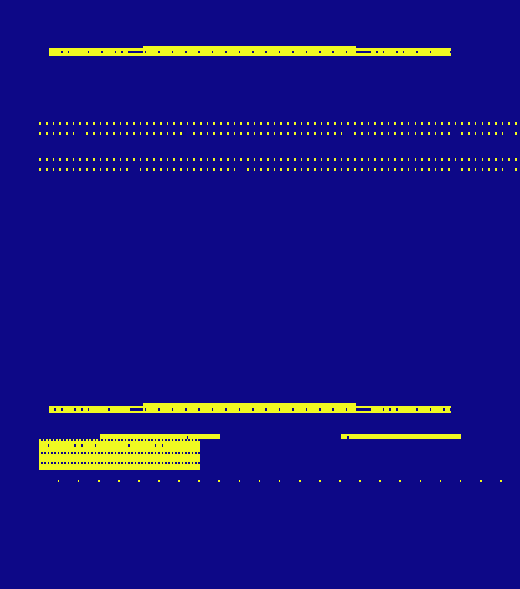}};
		
		  \node[draw, rectangle, fill=kit-blue30, minimum height=0.8cm, above right= 0.6cm and -0.2cm of img.north] (phy){ \begin{tabular}{c}
		       Physical  \\
		       Channel 
		  \end{tabular}};
		\node[draw, rectangle, fill=kit-green30, minimum width=1.8cm, minimum height=0.8cm, left=1cm of phy] (gnb) {gNB};

		\node[draw, rectangle, fill=kit-green30, minimum width=1.5cm, minimum height=0.8cm, above right=0.2cm and 1cm of phy.east] (ue1) {UE-1};
		\node[draw, rectangle, fill=kit-green30, minimum width=1.5cm, minimum height=0.8cm, below right=0.4cm and 1cm of phy.east] (ue2) {UE-N};
		\node[draw=none, fill=none, right=1.4cm of phy.east] (dots) {$\vdots$};
		
		\draw[<->, thick] (gnb.east) -- (phy.west);
		\draw[<->, thick] ([yshift=0.2cm]phy.east) -- (ue1.west);
		\draw[<->, thick] ([yshift=-0.2cm]phy.east) -- (ue2.west);

        \draw[-, gray] (phy.south west) -- (img.north west);
        \draw[-, gray] (phy.south east) -- (img.north east);
        
		\fill[kit-blue30, opacity=0.4] (img.south west) rectangle (img.north east);
		\foreach \i in {0,...,16} {
			\draw[gray,thin] ([xshift=-0.01cm,yshift={-0.06cm-0.31cm*\i}]img.north east) -- ++(0.2cm,0);
		}
		\foreach \i in {0,...,9} {
			\node at ([xshift=0.6cm,yshift={-0.525cm-0.31cm*\i}]img.north east) {{\footnotesize Slot \i}};
		}
		
		\draw[gray,thin] ([xshift=-0.01cm,yshift={-0.06cm-0.31cm*1}]img.north east) -- ++(1.3cm,0);
		\draw[gray,thin] ([xshift=-0.01cm,yshift={-0.06cm-0.31cm*6}]img.north east) -- ++(.4cm,0);
		\draw[gray,thin] ([xshift=-0.01cm,yshift={-0.06cm-0.31cm*11}]img.north east) -- ++(1.3cm,0);
		\draw[gray,thin] ([xshift=-0.01cm,yshift={-0.06cm-0.31cm*16}]img.north east) -- ++(.4cm,0);
		
		\node[rotate=90] (sfn) at ([xshift=1.7cm,yshift={-0.06cm-0.31cm*6}]img.north east) {\begin{tabular}{c}
				System Frame\\
				Duration = \SI{10}{ms}
		\end{tabular}};
		
		\node (PDCCH) at ([xshift=-1cm,yshift={-0.26cm-0.31cm*11}]img.north west) {{\small PDCCH}};
		\draw[->,kit-green70] ([xshift=-0.1cm]PDCCH.east) -- ++(.87cm,-0.11cm);
		\draw[->,kit-green70] ([xshift=-0.1cm,yshift=0.05cm]PDCCH.east) -- ++(2.14cm,-0.18cm);
		
		\node (PDSCH) at ([xshift=-1cm,yshift={-0.26cm-0.31cm*12}]img.north west) {{\small PDSCH}};
		\draw[->,kit-green70] ([xshift=-0.1cm]PDSCH.east) -- ++(.45cm,0.04cm);
		\draw [decorate, decoration={brace, amplitude=2.5pt}, thick, kit-green70,rotate=180]
		([xshift=-0.5cm,yshift=0.1cm]PDSCH.east) -- 
		([xshift=-0.5cm,yshift=-0.17cm]PDSCH.east);
		
		\node (SSB) at ([xshift=-1cm,yshift={-0.26cm-0.31cm*4}]img.north west) {{\small SSB}};
		\draw[->,kit-green70] ([yshift=-0.05cm]SSB.east) -- ++(.81cm,-2cm);
		\draw[->,kit-green70] ([yshift=0.05cm]SSB.east) -- ++(.81cm,1cm);

        \draw[->] (img.north west) -- ([yshift=-0.25cm]img.south west);
        \draw[->] (img.north west) -- ([xshift=0.3cm]img.north east);
        \node at ([xshift=0.3cm, yshift=2.5mm]img.north east) {$f$};
        \node at ([xshift=-0.2cm, yshift=-2mm]img.south west) {$t$};

	\end{tikzpicture}
	\caption{Exemplary system setup with one gNB and multiple UEs. The SSB is the first signal detected and allows for synchronization with the base station signal. PDCCH carries control information for the individual terminals, and the PDSCH carries data signals. Each system frame is divided into 10 slots, in this configuration, each of 1ms.}
	\label{fig:system-setup}
\end{figure}

In time domain, the 5G physical layer is structured in slots and system frames. A system frame is divided into 10 slots. In our case, a slot corresponds to \SI{1}{\ms}. System frames are identified by the \ac{SFN}, an integer number ranging from 0 to 1023. The physical spectrum and the slot structure can be seen in Figure \ref{fig:system-setup}.

\subsection{Transformer}
The machine learning architecture used in our research is called Transformer and was introduced by Vaswani et al. \cite{attention} in 2017. They rapidly outperformed traditional machine learning methods such as recurrent neural networks and \ac{LSTM} \cite{lin_survey_2021}, first in language processing and later in a broader field of applications. We chose this architecture because they showed remarkable results, for different lengths of input. For instance, when using ChatGPT, we don't have to worry about the length of our questions; within reasonable limits, it can handle varying input sizes without issue.

We utilize the implementation of Andrej Karpathy, known as nanoGPT \cite{nanogpt}, where we tuned the hyperparameter to fit our task.

\section{System Setup}
Our study is centered on forecasting protocol messages, and the goal is to predict the content of the next slot based on the content of the past 10 slots. The uplink and downlink are thereby combined. %

To test the system, different setups are simulated. In the first setup, two \acp{UE} are connected to a \ac{gNB}. Traffic is generated in downlink direction using the software iperf3 for both \acp{UE} using TCP and no speed limitation. Taking the same configurations into account, the number of \acp{UE} was modified between 2, 3, and 10. In the fourth configuration, the three traffic directions, downlink, uplink, and bi-directional, are each assigned to one \ac{UE}.

The \acp{UE} are connected to the \ac{gNB} with a perfect channel without noise. The recording duration is configured to approximately \SI{500} {\s}. Band n3 and a bandwidth of \SI{20}{\MHz} were utilized for all setups. 

\subsection{Environment}
For data generation, we rely on the open-source implementation srsRAN \cite{srsran2024} and GNU Radio for channel emulation. srsRAN provides a full-stack 5G implementation for \ac{gNB} and \ac{UE}. For the 5G core network, we utilize open5gs\cite{open5gs2024}, an open-source solution. 
Using this implementation, we are able to wiretap the communication between \ac{gNB} and \ac{UE} on physical layer providing access to all control signalling, which is not encrypted compared to user plane payload. We focus on the PHY layer as this information is accessible to a jammer as shown in the work of Falkenberg and Wietfeld \cite{Falkenberg2019a}. Using the control messages, one can extract which channel occupies which resources of the OFDM grid.
\begin{lstlisting}[style=myshell,caption=Exemplary control messages exchanged between gNB and UE (shortened version for a \ac{PDCCH} message)., label=list:log]
2024-11-26T08:33:52.600134 [PHY] [265.8] 
PDCCH: format=1_0 rnti=0x4601
PDSCH: rnti=0x4601 prb=[0, 96) symb=[1, 14)\end{lstlisting}

Listing \ref{list:log} shows the exemplary data used for our research. The first entries after the date of recording are the corresponding layer, and the \ac{SFN}, in this case, the physical layer (PHY) and \ac{SFN} 265. The slot number is the first decimal, i.e., 8. Afterwards, the different channel names are given; in this case, it is a downlink control message transmitted by the \ac{PDCCH} and the downlink data transmission over the \ac{PDSCH}. For channels with flexible resource assignment like \ac{PDSCH}, we need the subcarriers used and the occupied slots. The subcarriers are indicated by the \ac{PRB} number. The slots are given with the entry \textit{symb}. For example, both channels contain an entry for the \ac{RNTI}, indicating the target \ac{UE} of this message. The \ac{PDCCH} also has the \ac{DCI} indicator, given as \textit{format} in Listing \ref{list:log}. In this case, the \ac{DCI} is 1\us0, which indicates that in this frame the UE receives data through the \ac{PDSCH}.

\subsection{Data Preprocessing and Tokenization}
A Transformer needs the input to be represented in a set of tokens. All possible tokens used in our setup are listed in Listing \ref{list:tokens}. The tokenization represents the outcome of several development iterations. In total, there are 32 tokens. It starts with an empty token, which can be used when a slot contains no message. The hexadecimal digits from 0 to F serve multiple purposes, e.g., \ac{RNTI}, slot number, \ac{PRB}, and symbol numbering. A colon indicates the start of the slot and a semicolon the end. The comma is used to separate channels and to separate the start and end of the time-frequency intervals. We represent time resources enclosed by parentheses and frequency resources enclosed by square brackets. The last part of the tokens contains all the non-single-character tokens. In earlier experiments, we used \ac{PDCCH} without the information about the \ac{DCI}. Later, two new tokens were added to include this information. 

\begin{lstlisting}[style=myshell,caption=List of all tokens used to represent the physical layer messages exchanged., label=list:tokens]
"|\coloredtext[kit-green30]{}|", "|\coloredtext[kit-green30]{0}|", "|\coloredtext[kit-green30]{1}|", "|\coloredtext[kit-green30]{2}|", "|\coloredtext[kit-green30]{3}|", "|\coloredtext[kit-green30]{4}|", "|\coloredtext[kit-green30]{5}|", "|\coloredtext[kit-green30]{6}|",
"|\coloredtext[kit-green30]{7}|", "|\coloredtext[kit-green30]{8}|", "|\coloredtext[kit-green30]{9}|", "|\coloredtext[kit-green30]{A}|", "|\coloredtext[kit-green30]{B}|", "|\coloredtext[kit-green30]{C}|", "|\coloredtext[kit-green30]{D}|", "|\coloredtext[kit-green30]{E}|",
"|\coloredtext[kit-green30]{F}|", "|\coloredtext[kit-green30]{:}|", "|\coloredtext[kit-green30]{;}|", "|\coloredtext[kit-green30]{,$\phantom{!}\!\!\!$}|", "|\coloredtext[kit-green30]{/}|", "|\coloredtext[kit-green30]{[}|", "|\coloredtext[kit-green30]{]}|", "|\coloredtext[kit-green30]{(}|", 
"|\coloredtext[kit-green30]{)}|", "|\coloredtext[kit-green30]{PHY: PDCCH}|", "|\coloredtext[kit-green30]{PHY: PDSCH}|", "|\coloredtext[kit-green30]{PHY: PRACH}|",
"|\coloredtext[kit-green30]{PHY: PUCCH}|", "|\coloredtext[kit-green30]{PHY: PUSCH}|",
"|\coloredtext[kit-green30]{PHY: PDCCH - DCI:0\us0}|", "|\coloredtext[kit-green30]{PHY: PDCCH - DCI:1\us0}|"\end{lstlisting}

The result of tokenizing the control messages, as seen in Listing \ref{list:log}, is illustrated in Table \ref{tab:tokens}. Thereby, the \ac{SFN} is omitted, and only the slot is extracted. The \ac{RNTI} is shortened to two digits, and all decimal numbers in the messages are converted to hexadecimal for the tokens.
\vspace{-4mm}
\begin{table}[ht]
    \caption{Tokenization process}
    \begin{tabularx}{\columnwidth}{XX}
    \toprule
    Message entry & Tokens \\
    \midrule
    {[}PHY{]} {[}265.\textcolor{kit-green100}{8}{]} 
    & \textcolor{kit-green100}{8}:\\

    \textcolor{kit-green100}{PDCCH}: format=\textcolor{kit-green100}{1\us0} 
    & \textcolor{kit-green100}{PHY: PDCCH - DCI:1\us0}\\
    
    rnti=0x46\textcolor{kit-green100}{01} 
    & \textcolor{kit-green100}{01}\\

    & ,\\

    \textcolor{kit-green100}{PDSCH}: 
    & \textcolor{kit-green100}{PHY: PDSCH}\\

    rnti=0x46\textcolor{kit-green100}{01} 
    & \textcolor{kit-green100}{01}\\

    prb={[}\textcolor{kit-green100}{0}, \textcolor{kit-red100}{96})
    & {[}\textcolor{kit-green100}{0}, \textcolor{kit-red100}{60}{]}\\

    symb={[}\textcolor{kit-green100}{1}, \textcolor{kit-red100}{14})
    & (\textcolor{kit-green100}{1}, \textcolor{kit-red100}{E})\\

    & ;\\
    \bottomrule
    \end{tabularx}
    
    \label{tab:tokens}
\end{table}

\vspace{-4mm}

\section{Message Predictor}
\begin{figure}[t]
    \centering
    \begin{tikzpicture}[node distance=1cm, >=latex, thick]
        \node[draw, rectangle, minimum width=1.5cm, minimum height=0.5cm, fill=kit-green30] (sfn1) {{\small \begin{tabular}{l}
        		  {\footnotesize 8:PHY: PDCCH - DCI1\us0... }\\
        		{\footnotesize 9:PHY: PDSCH01[0,6]... }\\
        		{\footnotesize 0:PHY: PDSCH07[C,12]... }\\
        		{\footnotesize 1:PHY: PDCCH - DCI1\us0... }\\
        		{\footnotesize 2:PHY: PDCCH - DCI1\us0... }\\
        		{\footnotesize 3:PHY: PUCCH0A(0,E)... }\\
        		{\footnotesize 4:PHY: PDCCH - DCI1\us0... }\\
        		{\footnotesize 5:PHY: PDCCH - DCI0\us0... }\\
        		{\footnotesize 6:PHY: PDSCH02[6,C]... }\\
        		{\footnotesize 7:PHY: PDCCH - DCI1\us0...} 
        	\end{tabular}}};

        \node[draw, rectangle, minimum width=2.5cm, minimum height=1.3cm, fill=kit-blue30, below=0.6cm of sfn1] (Transformer) {%
            \begin{tabular}{c}
                \textbf{Message Prediction}\\
                \textbf{Transformer}
            \end{tabular}
        };

        \node[draw, rectangle, minimum width=1.5cm, minimum height=0.5cm, fill=kit-yellow30, below=0.6cm of Transformer] (sfn11) {{\footnotesize 8:PHY: PDCCH - DCI1\us0...}};

        \draw[->] (sfn1.south) -- (Transformer);

        \draw[->] (Transformer) -- (sfn11);

    \end{tikzpicture}
    \caption{The message predictor has approximately 10 slots as input and predicts the following slot. The number of slots in the input depends on the number of tokens per slot. The maximum number of input tokens is 1024. Sampling is repeated until the separator token is sampled.}
    \label{fig:flowgraph}
\end{figure}
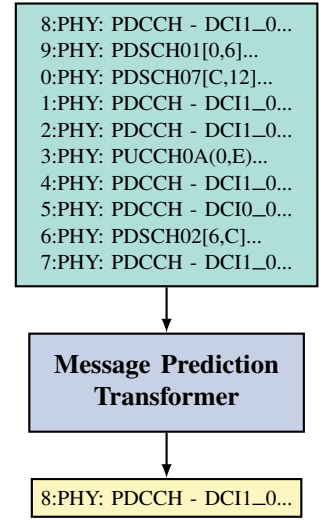

To predict the tokens of the next slot, we propose a Transformer-based setup, see Figure \ref{fig:flowgraph}. This Transformer has an input window length of 1024 tokens and predicts the 1025th token of the sequence. During validation, this procedure is repeated until the slot separator token is predicted. Using a slot separator allows for flexible token counts per slot.

Table \ref{tab:hyperparameter} lists the hyperparameters used for the Transformer-based predictor. These values stem from various experiments and may still allow for further optimization, which lies beyond the scope of this fundamental work.

\begin{table}[t]
    \caption{Hyperparameter for Transformer-based predictor}
    \begin{tabularx}{\columnwidth}{XX}
    \toprule
    Parameter & Value \\
    \midrule
    Input length & 1024 \\
    Embedding dimensions & 8 \\
    Layers  & 3 (Decoder structure)\\
    Multi-head attention & 8 \\
    Tokens & 32 \\
    Model parameter size & $\sim$ 2700\\
    Batch size & 2048 \\
    Training epochs & 4000 \\
    Training dataset size & $\sim$ \SI{250}{\kilo{\ tokens}} \\
    Validation dataset size & $\sim$ \SI{63}{\kilo{\ tokens}} \\
    \bottomrule
    \end{tabularx}
    
    \label{tab:hyperparameter}
\end{table}

\subsection{Syntax Checker}
The first training runs showed the vulnerability of the predictor to error propagation. Often, the sequence starts reasonably, while the later tokens suffer from more errors. Errors in the predicted channels can be acceptable, but when the slot number is already wrong, this introduces unnecessary errors. During evaluation, a syntax checker is added to reduce error propagation. 

First, the syntax checker compares the slot numbering, if it is counting up, with the exception that 0 follows 9. After the slot number, a colon needs to appear, and then the channels start. Each channel needs to be separated with a comma, and each channel’s message needs to start with the layer and channel description. After that, the \ac{RNTI} can follow. If a square bracket or bracket starts, it must be followed by two integer numbers separated by a comma and closed with the corresponding closing bracket.

\subsection{Validation Metrics - Levenshtein distance}
To get a similarity measure of sequences, the Levenshtein metric\cite{levenshtein} takes two sequences and compares what it takes to transform one into the other. Allowed operations are adding, removing, and changing entries.

\begin{figure}[ht!]
    \centering
    \begin{tikzpicture}[node distance=0.7cm, >=latex, thick]
        \node[draw, rectangle, minimum width=1.5cm, minimum height=0.6cm, fill=kit-green30] (input) {2 3 4 6 7 8};

        \node[draw, rectangle, minimum width=2.5cm, minimum height=1.5cm, fill=kit-blue30, right=of input] (edit) {\phantom{1} 2 3 4 6 7 8};

        \coordinate (crosscenter) at ([xshift=1.27cm,yshift=-0.75cm]edit.north west);
        \draw[kit-yellow30, thick] 
            ([xshift=-0.07cm,yshift=0.12cm]crosscenter) -- ([xshift=0.07cm,yshift=-0.12cm]crosscenter);
        \draw[kit-yellow30, thick] 
            ([xshift=0.07cm,yshift=0.12cm]crosscenter) -- ([xshift=-0.07cm,yshift=-0.12cm]crosscenter);
            
		\draw[->,color=kit-yellow30] ([yshift=0.35cm]crosscenter) -- ([yshift=0.12cm]crosscenter);
		\node at ([yshift=0.5cm]crosscenter) {5};
		
		\coordinate (crosscenter2) at ([xshift=2.15cm,yshift=-0.75cm]edit.north west);
		\draw[kit-red100, thick] 
		([xshift=-0.07cm,yshift=0.12cm]crosscenter2) -- ([xshift=0.07cm,yshift=-0.12cm]crosscenter2);
		\draw[kit-red100, thick] 
		([xshift=0.07cm,yshift=0.12cm]crosscenter2) -- ([xshift=-0.07cm,yshift=-0.12cm]crosscenter2);
		
		\coordinate (one) at ([xshift=0.38cm,yshift=-0.75cm]edit.north west);
		\draw[->,color=kit-green100] ([yshift=-0.37cm]one) -- ([yshift=-0.15cm]one);
		\node at ([yshift=-0.55cm]one) {1};
		\node[draw, rectangle, color=kit-green100, fill=kit-green100] at (one) {};

        \draw[->] (input.east) -- (edit.west);
   
        \node[draw, rectangle, minimum width=1.5cm, minimum height=0.6cm, fill=kit-yellow30, right=of edit] (output) {1 2 3 5 6 7};

        \draw[->] (edit.east) -- (output.west);

        \node[below=0.2cm of input] {\textbf{Sequence 1}};
        \node[below=0.01cm of edit] {\textbf{Edit operations}};
        \node[below=0.2cm of output] {\textbf{Sequence 2}};
    \end{tikzpicture}
    \caption{Two example sequences with the Levenshtein distance of three.}
    \label{fig:levenshtein_example}
\end{figure}
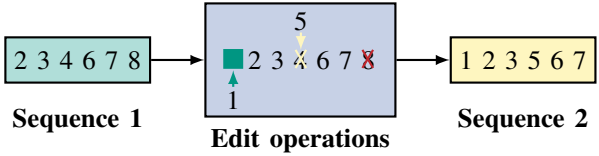

As an example, two sequences can be seen in Fig. \ref{fig:levenshtein_example}. To transform the first into the second, the following steps must be performed: add 1 at the beginning, change 4 to 5, and remove 8, i.e., the Levenshtein distance is three. 

The Levenshtein distance between the sequences $\mathbf{x}$ and $\mathbf{y}$ is denoted as,
$\text{L}(\mathbf{x},\mathbf{y})$,
where $\mathbf{y} \in \mathbb{N}^k$ and $\mathbf{x} \in \mathbb{N}^l$.

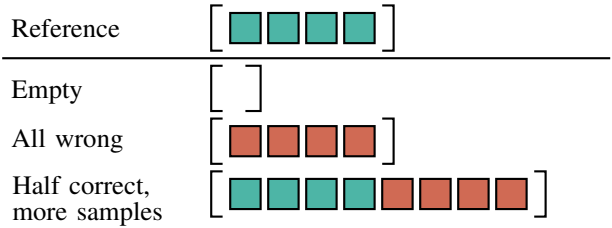
\begin{figure}[ht]
    \centering
    \begin{tikzpicture}[baseline]
        \node at (-3,0.2) [anchor=west] {Reference};
        \draw[thick] (-0.1,0.5) -- (-0.25,0.5) -- (-0.25,-0.1) -- (-0.1,-0.1);
        \draw[thick] (2,0.5) -- (2.15,0.5) -- (2.15,-0.1) -- (2,-0.1);

        \foreach \i in {0,...,3} {
            \fill[kit-green70] (\i*0.5,0) rectangle ++(0.4,0.4);
            \draw[thick] (\i*0.5,0) rectangle ++(0.4,0.4);
        }

        \draw[thick] (-3,-0.2) -- (5,-0.2);

        \node at (-3,-0.65) [anchor=west] {Empty};
        \draw[thick] (-0.1,-0.3) -- (-0.25,-0.3) -- (-0.25,-.9) -- (-0.1,-.9);
        \draw[thick] (0.2,-0.3) -- (0.4,-0.3) -- (0.4,-.9) -- (0.2,-.9);

        \node at (-3,-1.3) [anchor=west] {All wrong};
        \draw[thick] (-0.1,-1) -- (-0.25,-1) -- (-0.25,-1.6) -- (-0.1,-1.6);
        \draw[thick] (2,-1) -- (2.15,-1) -- (2.15,-1.6) -- (2,-1.6);
        \foreach \i in {0,...,3} {
            \fill[kit-red70] (\i*0.5,-1.5) rectangle ++(0.4,0.4);
            \draw[thick] (\i*0.5,-1.5) rectangle ++(0.4,0.4);
        }

        \node at (-3.2,-2.1) [anchor=west] {\renewcommand{\arraystretch}{0.8}\begin{tabular}{l}Half correct,\\ more samples\end{tabular}};
        \draw[thick] (-0.1,-1.7) -- (-0.25,-1.7) -- (-0.25,-2.3) -- (-0.1,-2.3);
        \draw[thick] (4,-1.7) -- (4.15,-1.7) -- (4.15,-2.3) -- (4,-2.3);
        \foreach \i in {0,...,3} {
            \fill[kit-green70] (\i*0.5,-2.2) rectangle ++(0.4,0.4);
            \draw[thick] (\i*0.5,-2.2) rectangle ++(0.4,0.4);
        }
        \foreach \i in {4,...,7} {
            \fill[kit-red70] (\i*0.5,-2.2) rectangle ++(0.4,0.4);
            \draw[thick] (\i*0.5,-2.2) rectangle ++(0.4,0.4);
        }

    \end{tikzpicture}
    \caption{Three different sequences with the same \ac{RL} distance of 1.}
    \label{fig:rl1}
\end{figure}

One flaw of the Levenshtein distance arises when comparing two reference sequences $\mathbf{y_1} \in \mathbb{N}^{k_1}$ and $\mathbf{y_2}\in \mathbb{N}^{k_2}$ where $k_1 > k_2$ with  $\mathbf{x_1} \in \mathbb{N}^{l_1}$ and $\mathbf{x_2} \in \mathbb{N}^{l_2}$ respectively. Provided that the Levenshtein distances for these cases are equal, i.e., 
$\text{L}(\mathbf{x_1},\mathbf{y_1}) = \text{L}(\mathbf{x_2},\mathbf{y_2})$, it follows that $\mathbf{x_1}$ and $\mathbf{y_1}$ have a higher number of matching tokens compared to the tokens shared between $\mathbf{x_2}$ and $\mathbf{y_2}$. However, the Levenshtein distance metric does not convey this information. Therefore, we apply the \ac{RL} distance.
\newtheorem{definition}{Definition}
\begin{definition}[Relative Levenshtein]
$$\text{RL}(\mathbf{x}, \mathbf{y}): = \frac{\text{L}(\mathbf{x},\mathbf{y})}{k}.$$
This measure returns the percentage of wrong tokens, compared to the total length $k$ of the reference sequence $\mathbf{y}$. 
\end{definition}
A relative Levenshtein distance of 0 means a correct prediction. Other results of the \ac{RL} don't directly indicate the type of error. Figure \ref{fig:rl1} depicts a scenario in which the \ac{RL} distance is always equal to 1 with different sequences.

\subsection{Validation Metrics - Precision}
In addition to the Levenshtein distance, we can also investigate the precision of tokens and fields in the prediction. The control messages can be denoted as a sequence of tokens $\mathcal{T} = \{ t_0, \dots, t_n \}$. Additionally, we define a subset $\mathcal{C} = \{c_0, \dots, c_n\}$, where $c_i$ is one of the channel tokens. During evaluation, we extract all channel tokens that appear and store them in $\mathcal{C}^\text{val} = \{c^\text{val}_0, \dots, c^\text{val}_n\}$. Then $\mathrm{P}(c_i) = \mathrm{P}(c_i^\text{val})$ gives the frequency in percent of this token. To measure how often the token was predicted correctly, we use the \ac{TP} function 
\begin{align*}
    \text{TP}(c_i) &= \mathrm{P}(c_i \in \mathcal{C}^\text{val} \, | \, c_i \in \mathcal{C}^\text{pred}) \\  &=\mathrm{E} \mkern-4mu \left( \frac{|c_i \in \mathcal{C}^\text{val} \wedge c_i \in \mathcal{C}^\text{pred}|}{|c_i \in \mathcal{C}^\text{pred}|}\right),
\end{align*}
where $\mathcal{C}^\text{val}$ denotes the validation sequence and $\mathcal{C}^\text{pred}$ the predicted sequence. If the channel was predicted correctly, then we can also evaluate the precision of the \ac{RNTI} denoted as $r$. The \ac{RNTI} true positive function is defined as 

$$\text{TP}(r|c_i) = \mathrm{P}(r \in \mathcal{T}_\text{val}\, | \, r \in \mathcal{T}_\text{pred} \wedge c_i \in \mathcal{C}_\text{val} \wedge c_i \in \mathcal{C}_\text{pred})$$
$$=\mathrm{E} \mkern-4mu \left( \frac{|r \in \mathcal{T}_\text{val} \wedge r \in \mathcal{T}_\text{pred} \wedge c_i \in \mathcal{C}_\text{val} \wedge c_i \in \mathcal{C}_\text{pred}|}{|c_i \in \mathcal{C}^\text{pred}|}\right),
$$
where $r$ represents the tokens that resemble the \ac{RNTI}. We conducted additional evaluations, but for brevity, we have excluded them from this paper.

\section{Results}
We focus our discussion on the results obtained with the syntax checker, as they outperform those without. Nevertheless, some figures report both configurations for comparison. First, we present the results of the data set using 2 \acp{UE} with data received in downlink. From the validation set, 11 consecutive slots are chosen randomly. The initial 10 slots act as the input for the predictor, while the 11th serves as reference for evaluating the prediction. One such reference can be seen in Table \ref{tab:tokens}. The sampled tokens for the same input can be seen in Listing \ref{list:outres}. The alternating colors represent the borders of the tokens.
\begin{lstlisting}[style=myshell,caption=Predicted tokens for slot number 8 with same 10 previous slots inputed as in Table \ref{tab:tokens}., label=list:outres]
|\coloredtext[kit-green30]{8}||\coloredtext[kit-red30]{:}||\coloredtext[kit-green30]{PHY: PDCCH - DCI:1\us0}||\coloredtext[kit-red30]{0}||\coloredtext[kit-green30]{1}||\coloredtext[kit-red30]{\!,$\phantom{!}\!\!$}||\coloredtext[kit-green30]{PHY:PDSCH}||\coloredtext[kit-red30]{0}||\coloredtext[kit-green30]{1}||\coloredtext[kit-red30]{[}||\coloredtext[kit-green30]{0}||\coloredtext[kit-red30]{\!,$\phantom{!}\!\!$}||\coloredtext[kit-green30]{6}||\coloredtext[kit-red30]{0}||\coloredtext[kit-green30]{]}|
|\coloredtext[kit-red30]{(}||\coloredtext[kit-green30]{1}||\coloredtext[kit-red30]{\!,$\phantom{!}\!\!$}||\coloredtext[kit-green30]{E}||\coloredtext[kit-red30]{)}||\coloredtext[kit-green30]{\!,$\phantom{!}\!\!$}||\coloredtext[kit-red30]{PHY: PUCCH}||\coloredtext[kit-green30]{0}||\coloredtext[kit-red30]{1}||\coloredtext[kit-green30]{(}||\coloredtext[kit-red30]{0}||\coloredtext[kit-green30]{\!,$\phantom{!}\!\!$}||\coloredtext[kit-red30]{E}||\coloredtext[kit-green30]{)}||\coloredtext[kit-red30]{;}|
\end{lstlisting}
The initial segment is accurately predicted, indicating that both the syntax and content can be effectively learned by the Transformer. The last channel, the \ac{PUCCH}, is not in the reference output and is added wrongly by the Transformer. Still, the syntax seems reasonable, but not all added tokens are needed in this case, i.e., the Levenshtein distance is 9.

\begin{figure}[ht!]
	\centering
	\begin{tikzpicture}
		\begin{axis}[view={0}{90},
            width=70mm,
			xlabel=Transformer Output - Token,
            ylabel=Sampling step,
            xticklabel style={xshift=0.1em},
            ytick={0,9,19,29,39},
            yticklabels={1,10,20,30,40},  
            tick align=outside,
            tick style={black, thick},  
            ytick pos=left, 
            xtick pos=left,      
            y dir=reverse,
            colorbar,
			colorbar style={
                ylabel=Token probability,
			}]
            \addplot3[
                scatter, 
                mesh, 
                mark=square*
            ] 
            table [
                x=token, 
                y=sample, 
                z=prob, 
                col sep=comma
            ] {fig/probs.csv};
            \pgfplotsset{axis equal image}
		\end{axis}
	\end{tikzpicture}
	\caption{For each sample, the output probabilities of the Transformer are shown. The original input contains 10 slots, and the 11th is predicted. There are 30 tokens drawn until the end token is sampled in the last step.}
	\label{fig:probscoolwarm}
\end{figure}
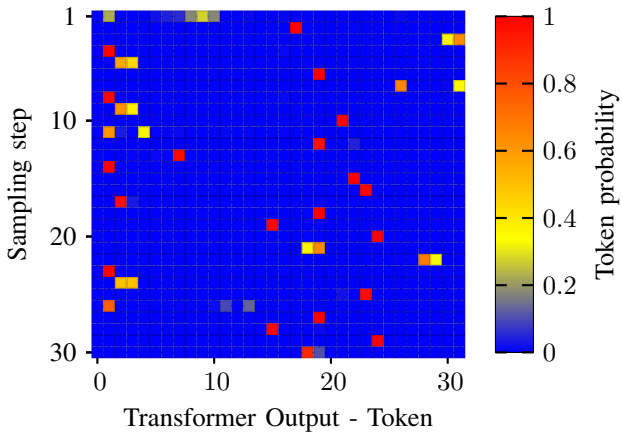

The output probabilities for the consecutive tokens drawn from the Transformer can be seen in Fig. \ref{fig:probscoolwarm}.

The color indicates the probability that the Transformer assigns to each token. Typically, the probability is close to 1, suggesting that tokens are very predictable, with only a single option being evident. In some cases, more options are possible, for example, in sample 21, two options,
``comma" and ``semicolon" are possible. Here, the ``comma" carries a greater probability, however, if the Transformer opted for the second alternative, the result displayed in Listing \ref{list:outres} would be correct.

\subsection{Validation with multiple UEs}
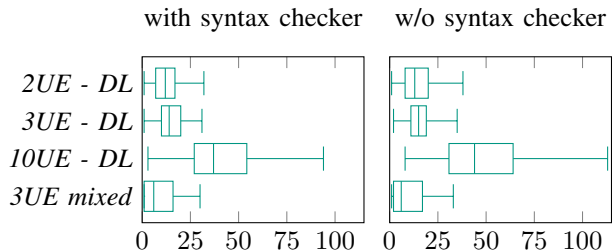
\begin{figure}[ht!]
	\begin{center}
		\begin{tikzpicture}
			\begin{axis}
				[
                title= with syntax checker,
				ytick={1,2,3,4},
				yticklabels={\textit{2UE - DL}, \textit{3UE - DL}, \textit{10UE - DL}, \textit{3UE mixed}},
                ytick style={draw=none}, 
				y dir=reverse,
				xtick={0,25,50,75,100},
				xmin=0,
				xmax=115,
				width=4.5cm,
				height=5cm,
				ymin=0.3,
				ymax=4.7,
				y=0.5cm,
				]
				\addplot+[kit-green100,boxplot prepared={
					median=12.0,
					upper quartile=17.0,
					lower quartile=7.0,
					upper whisker=32.0,
					lower whisker=1.0
				}] coordinates {};
				\addplot+[kit-green100,boxplot prepared={
					  median=14.0,
                    upper quartile=20.0,
                    lower quartile=10.0,
                    upper whisker=31.0,
                    lower whisker=1.0
				}] coordinates {};
				\addplot+[kit-green100,boxplot prepared={
                    median=37.0,
					upper quartile=54.25,
					lower quartile=27.0,
					upper whisker=94.0,
					lower whisker=3.0
				}] coordinates {};
				\addplot+[kit-green100,boxplot prepared={
					median=6.0,
					upper quartile=16.0,
					lower quartile=1.0,
					upper whisker=30.0,
					lower whisker=1.0
				}] coordinates {};
				
			\end{axis}
		\end{tikzpicture}\begin{tikzpicture}
			\begin{axis}
				[
                title=w/o syntax checker,
				ytick={1,2,3,4},
				yticklabels={},
                ytick style={draw=none}, 
				y dir=reverse,
				xtick={0,25,50,75,100},
				xmin=0,
				xmax=115,
				width=4.5cm,
				height=5cm,
				ymin=0.3,
				ymax=4.7,
				y=0.5cm,
				]
				\addplot+[kit-green100,boxplot prepared={
					median=13.0,
                    upper quartile=20.0,
                    lower quartile=8.0,
                    upper whisker=38.0,
                    lower whisker=1.0
                }] coordinates {};
				\addplot+[kit-green100,boxplot prepared={
                    median=15.0,
                    upper quartile=19.0,
                    lower quartile=11.0,
                    upper whisker=35.0,
                    lower whisker=2.0
				}] coordinates {};
				\addplot+[kit-green100,boxplot prepared={
                    median=44.0,
                    upper quartile=64.0,
                    lower quartile=30.75,
                    upper whisker=113.0,
                    lower whisker=8.0
				}] coordinates {};
				\addplot+[kit-green100,boxplot prepared={
                    median=6.0,
                    upper quartile=17.0,
                    lower quartile=2.0,
                    upper whisker=33.0,
                    lower whisker=1.0
				}] coordinates {};
				
			\end{axis}
		\end{tikzpicture}
	\end{center}
    \caption{Levenshtein distance for the different scenarios}
    \label{fig:boxld}
\end{figure}

First, we discuss the Levenshtein distance and the relative Levenshtein distance. When more users are added to the system, the complexity increases, and as expected, the Levenshtein distance increases, as can be seen in the box plots in Fig. \ref{fig:boxld}. The \textit{2UE - DL}, \textit{3UE - DL}, and \textit{10UE - DL} scenarios are using the same setup, but with different numbers of \acp{UE}. 

 The \textit{3UE mixed} setup differs from the previous setups, because each \ac{UE} has a different traffic pattern. So far, we cannot definitively explain the reason for the improved performance. A possible explanation is that the scheduler used is more predictable in this setup. In this scenario, the channels are scheduled more regularly, which has a substantial effect on the Levenshtein distance. If a channel is predicted that does not actually appear in the validation data, this leads to several incorrect tokens. Conversely, if the predicted channel is correct, then only its parameters may be wrong, which results in a smaller Levenshtein distance.
 
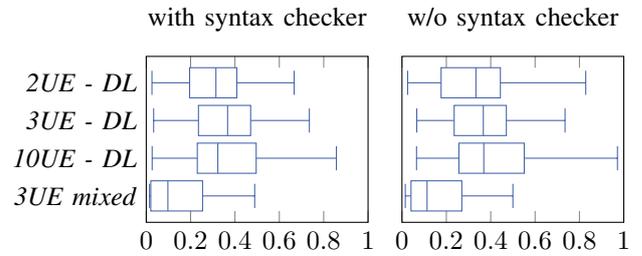
\begin{figure}[t]
	\begin{center}
		\begin{tikzpicture}
			\begin{axis}
				[
                title=with syntax checker,
				ytick={1,2,3,4},
				yticklabels={\textit{2UE - DL}, \textit{3UE - DL}, \textit{10UE - DL}, \textit{3UE mixed}},
                ytick style={draw=none}, 
				y dir=reverse,
				xtick={0,0.2,0.4,0.6,0.8,1},
				xmin=0,
				xmax=1,
				width=4.5cm,
				height=5cm,
				ymin=0.3,
				ymax=4.7,
				y=0.5cm,
				]
				\addplot+[kit-blue100,boxplot prepared={
                    median=0.31414473684210525,
                    upper quartile=0.40839517625231914,
                    lower quartile=0.19545454545454546,
                    upper whisker=0.6666666666666666,
                    lower whisker=0.02564102564102564
                }] coordinates {};
                \addplot+[kit-blue100,boxplot prepared={
                    median=0.36666666666666664,
                    upper quartile=0.47058823529411764,
                    lower quartile=0.23529411764705882,
                    upper whisker=0.7352941176470589,
                    lower whisker=0.03333333333333333
                }] coordinates {};
                \addplot+[kit-blue100,boxplot prepared={
                    median=0.32271485345255835,
                    upper quartile=0.49557522123893805,
                    lower quartile=0.23008849557522124,
                    upper whisker=0.8571428571428571,
                    lower whisker=0.02654867256637168
                }] coordinates {};
                \addplot+[kit-blue100,boxplot prepared={
                    median=0.09748854573222468,
                    upper quartile=0.25396825396825395,
                    lower quartile=0.02040816326530612,
                    upper whisker=0.4888888888888889,
                    lower whisker=0.014925373134328358
                }] coordinates {};
			\end{axis}
		\end{tikzpicture}\begin{tikzpicture}
			\begin{axis}
				[
                title=w/o syntax checker,
                yticklabels={},
				ytick={1,2,3,4},
                ytick style={draw=none}, 
				y dir=reverse,
				xtick={0,0.2,0.4,0.6,0.8,1},
				xmin=0,
				xmax=1,
				width=4.5cm,
				height=5cm,
				ymin=0.3,
				ymax=4.7,
				y=0.5cm,
				]
				\addplot+[kit-blue100,boxplot prepared={
                    median=0.3333333333333333,
                    upper quartile=0.44364024679639297,
                    lower quartile=0.17647058823529413,
                    upper whisker=0.8285714285714286,
                    lower whisker=0.02564102564102564
                }] coordinates {};
                \addplot+[kit-blue100,boxplot prepared={
                    median=0.36666666666666664,
                    upper quartile=0.47058823529411764,
                    lower quartile=0.23529411764705882,
                    upper whisker=0.7352941176470589,
                    lower whisker=0.06666666666666667
                }] coordinates {};
                \addplot+[kit-blue100,boxplot prepared={
                    median=0.36885245901639346,
                    upper quartile=0.5512123373150295,
                    lower quartile=0.25663716814159293,
                    upper whisker=0.9714285714285714,
                    lower whisker=0.06557377049180328
                }] coordinates {};
                \addplot+[kit-blue100,boxplot prepared={
                    median=0.11320754716981132,
                    upper quartile=0.2698412698412698,
                    lower quartile=0.04081632653061224,
                    upper whisker=0.5,
                    lower whisker=0.014925373134328358
                }] coordinates {};
			\end{axis}
		\end{tikzpicture}
	\end{center}
    \caption{Relative Levenshtein distance for the different scenarios}
    \label{fig:boxrld}
\end{figure}

 The increase of the Levenshtein distance for the increasing number of \acp{UE} is not only expected because of the growing complexity of the system, but also because each slot contains more messages. This leads to an increase in tokens per slot. Taking a look at the relative Levenshtein distance results in Fig. \ref{fig:boxrld}, we see that all three scenarios with the same traffic direction but an increasing number of users perform nearly equally on average. Only the outliers are getting worse with the increasing number of \acp{UE}.

Table \ref{tab:precision} presents the precision results for two of the four scenarios. We don't show the 2 and 3 UE DL scenarios because they perform similarly to the 10 UE scenario. 

Each channel token is given with its relative frequency $\mathrm{P}(c_i)$ and how often it is predicted correctly through $\mathrm{TP}(c_i)$. Additionally, when predicted correctly, we determined how often the \ac{RNTI} is also predicted correctly, denoted as $\mathrm{TP}(r|c_i)$.

\begin{table}[ht!]
    \caption{Frequency and precision of channels}
    \label{tab:precision}
    \centering
    \begin{tabularx}{\columnwidth}{llcCC}
    \toprule
    & Token & $\mathrm{P}(c_i)$ & $\mathrm{TP}(c_i)$ & $\mathrm{TP}(r|c_i)$\\
    \midrule
    \multirow{5}{*}{\rotatebox{90}{10 UEs}}
    & PDSCH & 28.90\% & 77.46\% & 39.69\%\\
    & PDCCH - DL & 31.90\% & 85.71\% & 61.38\%\\
    & PUCCH & 36.69\% & 85.27\% & 22.98\%\\
    & PUSCH & 1.10\% & 8.16\% & 0.00\%\\
    & PDCCH - UL & 1.42\% & 0.00\% & 0.00\%\\
    \midrule
    \multirow{5}{*}{\rotatebox{90}{3 UEs mixed}}
    & PDSCH & 20.47\% & 99.19\% & 91.62\%\\
    & PDCCH - DL & 20.47\% & 99.19\% & 92.97\%\\
    & PUCCH & 14.22\% & 85.00\% & 73.33\%\\
    & PUSCH & 27.94\% & 100.00\% & 86.67\%\\
    & PDCCH - UL & 16.90\% & 98.70\% & 82.08\%\\
    \bottomrule
    \end{tabularx}
    
\end{table}

By adding more \acp{UE}, the Transformer has proven to scale well with the complexity. However, with each added \ac{UE}, the mean number of tokens per slot increases. This implies that the number of slots represented in the input is inversely proportional to the number of \acp{UE}. The model can be changed to accommodate a longer window of input tokens.

For now, a new Transformer is trained on each scenario, and no cross-evaluation is performed. This should be addressed in further research.

\section{Conclusion}
In this work, we proposed a Transformer-based model to predict frames of a 5G communication system. Our experiments reveal that the syntax and latent states of the system are learnable by a Transformer model. The proposed method of representing the content of a slot yields promising results. The system performs well when accommodating messages of differing lengths across slots. The number of \acp{UE} can be varied without a significant decrease in the performance of the prediction.

\end{document}